\documentclass[10pt,journal,compsoc]{IEEEtran}
\IEEEoverridecommandlockouts
\AtBeginDocument{%
  \providecommand\BibTeX{{%
    \normalfont B\kern-0.5em{\scshape i\kern-0.25em b}\kern-0.8em\TeX}}}

\usepackage{multicol}
\usepackage{multirow}
\usepackage{amsmath,amsfonts}
\usepackage{amssymb}
\usepackage{array}
\usepackage{ragged2e}
\usepackage{textcomp}
\usepackage{stfloats}
\usepackage{url}
\usepackage{verbatim}
\usepackage{graphicx}
\usepackage{cite}
\usepackage{pifont}
\usepackage[ruled,vlined]{algorithm2e}
\usepackage[justification=centering]{caption}
\usepackage{enumitem}
\usepackage{subcaption}
\usepackage{booktabs}
\usepackage{makecell}
\usepackage{tcolorbox}
\usepackage{soul}
\usepackage{ragged2e}
\usepackage{threeparttable}
\usepackage{balance}
\usepackage{colortbl}

\newcommand{\toolname}{CasModaTest\xspace}

\newcommand{\phead}[1]{\vspace{1mm} \noindent {\bf #1}}

\newcommand{\intuition}[1]{
\begin{tcolorbox}[colback=white,boxrule=1pt,top=0pt,bottom=0pt,left=1pt,right=2pt,top=2pt,bottom=2pt]
\em #1
\end{tcolorbox}
}

\begin{document}
\title{\toolname: A Cascaded and Model-agnostic Self-directed Framework for Unit Test Generation}


\author{
Chao Ni, Xiaoya Wang, Liushan Chen, Dehai Zhao, Zhengong Cai,  Shaohua Wang, Xiaohu Yang
\IEEEcompsocitemizethanks{
\IEEEcompsocthanksitem Chao Ni, Xiaoya Wang, Zhengong Cai, and Xiaohu Yang are with the State Key Laboratory of Blockchain and Data Security, Zhejiang University, Hangzhou, China. 
Chao Ni is also with Hangzhou High-Tech Zone (Binjiang) Blockchain and Data Security Research Institute, Hangzhou, China.
E-mail: \{chaoni, wxy\_2000, cstcaizg, yangxh\}@zju.edu.cn
\IEEEcompsocthanksitem Liushan Chen is with ByteDance, China. E-mail: chenliushan@bytedance.com
\IEEEcompsocthanksitem Dehai Zhao is with Data61, Australia. E-mail: dehai.zhao@data61.csiro.au
\IEEEcompsocthanksitem Shaohua Wang is with Central University of Finance and Economics, China. E-mail: davidshwang@ieee.org
}
\thanks{Both Chao Ni and Xiaoya Wang contributed equally to this research. Zhengong Cai is the corresponding author}
}

\IEEEtitleabstractindextext
{%
\begin{abstract}
\justifying{
Though many machine learning (ML)-based unit testing generation approaches have been proposed and indeed achieved remarkable
performance, they still have several limitations in effectiveness and practical usage. 
More precisely, existing ML-based approaches (1) generate partial content of a unit test, mainly focusing on test oracle generation;
(2) mismatch the test prefix with the test oracle semantically; and (3) are highly bound with the close-sourced model, eventually damaging data security.
\\
In this paper, we propose a novel approach named \toolname, a cascaded, model-agnostic, and end-to-end unit test generation framework, to alleviate the above limitations.
Specifically, \toolname first splits the unit test generation task as two cascaded ones: test prefix generation and test oracle generation.
Then, to better stimulate models' learning ability, we manually build large-scale demo pools to provide \toolname with high-quality test prefixes and test oracles examples.
Finally, \toolname automatically assembles the generated test prefixes and test oracles and compiles or executes them to check their effectiveness, optionally appending with several attempts to fix the errors occurring in compiling and executing phases.
\\
To evaluate the effectiveness of \toolname, we conduct large-scale experiments on a widely used dataset (Defects4J) and compare it with four state-of-the-art (SOTA) approaches by considering two performance measures. 
The experimental results indicate that \toolname outperforms all SOTAs with a substantial improvement (i.e., 60.62\%-352.55\% in terms of accuracy, 2.83\%-87.27\% in terms of focal method coverage).
Besides, we also conduct experiments of \toolname on different open-source LLMs and find that \toolname can also achieve significant improvements over SOTAs (39.82\%-293.96\% and 9.25\%-98.95\% in
terms of accuracy and focal method coverage, respectively) in end-to-end unit test generation.
}
\end{abstract}

}



\maketitle

\section{Introduction}

Unit testing plays a critical role in software maintenance to assure the quality of software systems, which helps developers identify defects and errors as early as possible in the development process, reducing the overall cost of the product and hence improving developers' productivity~\cite{garousi2013survey,lee2012survey,barr2014oracle}.
However, manually writing high-quality unit tests is a difficult and time-consuming task.
Thus, various automated test case generation approaches have been proposed and these approaches can be divided into two categories: traditional unit test generation~\cite{pacheco2007randoop,fraser2011evosuite,2021tackletest} and machine learning-based unit test generation~\cite{tufano2020unit,alagarsamy2023a3test,xie2023chatunitest,schafer2023adaptive}.
These approaches have achieved promising results and indeed make progress in unit test generation scenarios, especially the machine learning-based approaches that have attracted much attention from academia to industry.
However, machine learning-based approaches still have several limitations in previous studies including (1) \ul{Impractical Usage}, (2) \ul{Unpaired Prefix and Oracle}, and (3) \ul{Close-sourced Binded Model}.
More precisely, a complete unit test case should consist of two parts: test prefix and test oracle. 
In practice, a developer or a tester indeed needs a complete unit to help them verify the quality of software.
However, currently,  most machine learning-based approaches focus on a part of unit test, such as generating test oracle only with test prefix given~\cite{sun2023revisiting,zhang2023saga,dinella2022toga}, which is not suitable for practical usage since the participants still need to carefully prepare the partial unit test.
Meanwhile, there are two types of unit tests~\cite{dinella2022toga}: tests with assertion oracles and tests with expected exception oracles. 
The former verifies the correctness of the return behavior, although they fail if any exception occurs, while the latter verifies whether the execution of the test prefix with invalid usage can raise a particular exception.
Therefore, the test prefix should be well semantically connected with the corresponding test oracle.
For example, the test prefix with the semantics of assertion should be assembled with the assertion oracle, the same applies to the test prefix that embodies the semantics of expected exception.
However, though several approaches~\cite{tufano2020unit,alagarsamy2023a3test,moradi2023effective} can generate a complete unit test, they may not well consider the connection between the test prefix and test oracle and confuse the two parts, which leads to an incorrect unit test.
Finally, with the demonstrated powerful capability of large language models (e.g., 
ChatGPT~\cite{openai2022chatgpt}), researchers have proposed approaches for unit test generation based on LLM but such approaches are generally relying on closed-sourced models' capabilities (e.g., ChatGPT-based approach ChatUnitTest~\cite{xie2023chatunitest}).

To address the aforementioned limitations, we proposed an automatic, cascaded, and LLM-agnostic unit test generation framework, \toolname, with two phases: the generation phase and the verification phase. 
The former phase generates a test prefix for the focal method and a corresponding semantic-related test oracle.
In the latter phase, the test prefix and the test oracle will be combined as a complete unit test for verification and following that, an automatic self-debug process will be triggered for addressing compile or execution errors with limited interactions with LLMs (e.g., ChatGPT or CodeLlama).
To achieve better unit tests by adopting LLM's in-context learning ability, we manually built two demonstration pools based on the widely used Defect4J dataset~\cite{just2014defects4j} for generating test prefixes and test oracles separately.
We evaluate the effectiveness of \toolname by comparing with three state-of-the-art deep learning-based (i.e., AthenaTest~\cite{tufano2020unit}, A3Test~\cite{alagarsamy2023a3test}) or ChatGPT-based approaches (i.e., ChatUnitTest~\cite{xie2023chatunitest}) with two widely used performance metrics (i.e., accuracy and focal method coverage).
The experimental results indicate the priority of \toolname.
More precisely, \toolname equipped with closed source LLM (i.e., gpt-3.5-turbo) significantly improves SOTAs by 97.19\%-352.55\%  and  2.83\%-87.27\%  in terms of accuracy and focal method coverage, respectively.
\toolname equipped with open source LLM (i.e., DeepSeek, Phind-CodeLlama) also obtains a great improvement over SOTAs by 71.66\%-293.96\% and 9.25\%-98.95\% in terms of accuracy and focal method coverage, respectively.
Eventually, our contributions are listed as follows:
\begin{itemize}[leftmargin=*]
    \item \textbf{A. Novel Self-directed LLM-based Unit Test Generation:} \toolname advances LLM-agnostic unit test generation.
We show that the two-stage and LLM-agnostic unit test generation framework can achieve better results over existing unit test generation directions.
\item  \textbf{B. Structured Demonstration Pool:} We manually build two types of demonstration pools for further prompting the two-stage unit test generation with the in-context learning ability of LLM.

\item  \textbf{C. Extensive Empirical Evaluation:} 
We evaluate \toolname against current SOTA deep learning-based and LLM-based tools on the widely studied Defects4J~\cite{just2014defects4j}.
\toolname outperforms the existing SOTA unit test generation approaches.
\end{itemize}

\label{sec:introduction}

\section{Motivation Example}

\begin{figure*}[htbp]
    \centering
    \includegraphics[width=.9\linewidth]{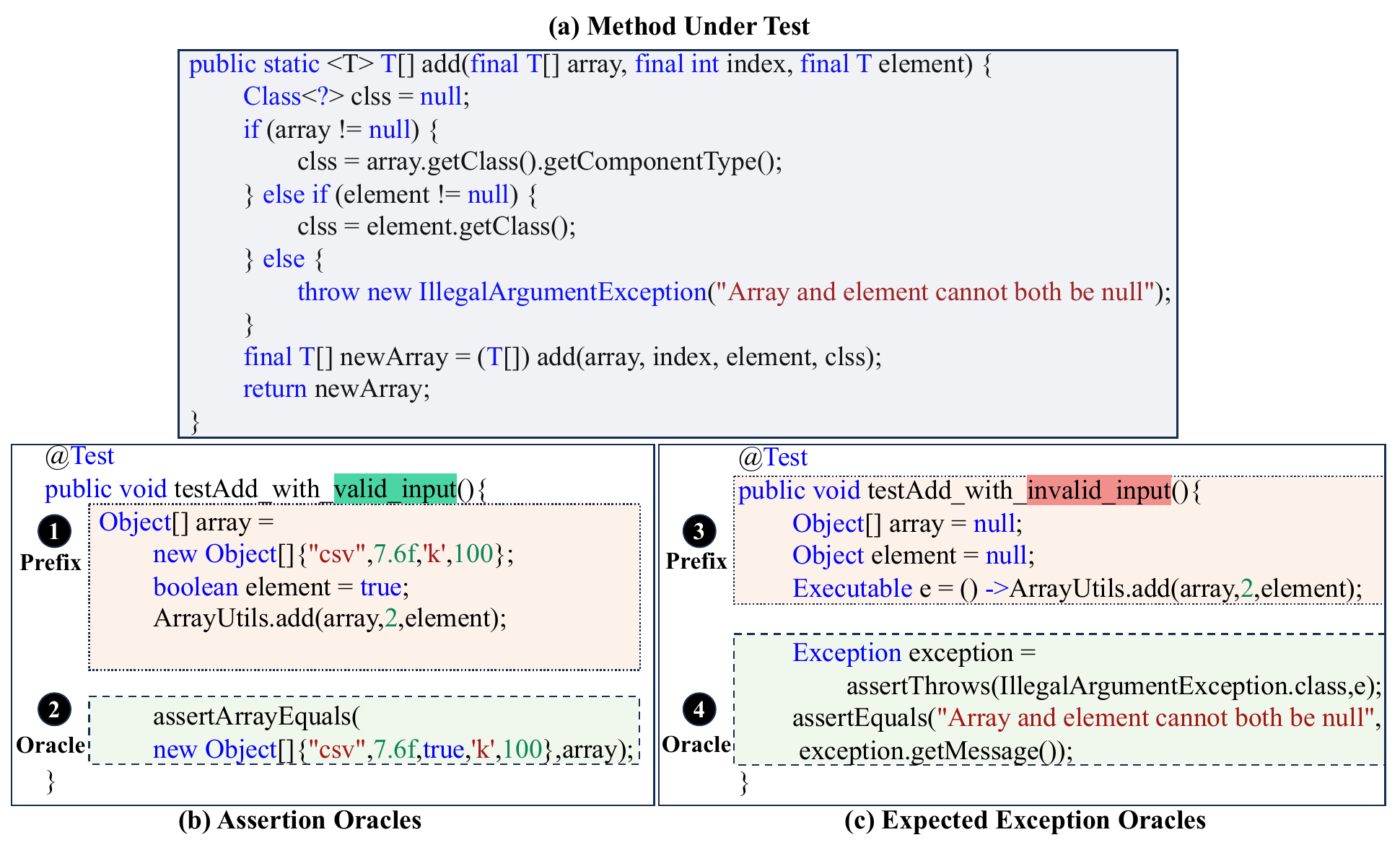}
\caption{(a) the method named \texttt{add} in class \texttt{ArrayUnits} from Apache Commons Lang~\cite{apachecommonlange} ; (b) an assertion oracle unit test to verify correct return behavior; (c) an expected exception oracle to verify that executing the test prefix with some invalid usage raises an exception. }
    \label{fig:motivation}
\end{figure*}

Fig.~\ref{fig:motivation} shows a method named \texttt{add} in the class \textit{ArrayUnits} from the Apache Commons Lang project~\cite{apachecommonlange} and its two unit test cases.
The function \textit{add} is to add an element to an array at a particular position (i.e., \textit{index}). 
Particularly, before adding an element to the array, it is necessary to check that both the element's value and the array itself are not \textit{NULL}.
Meanwhile, the given index should be inside the array's range.
If the given element and array are not \textit{NULL}, this method inserts the element into the target array at the given position. 
Therefore, there are two types of unit tests: tests with assertion oracles and tests with expected exception oracles, defined by previous work~\cite{dinella2022toga} by observing almost 200K developer-written tests. 
The \textbf{Test with Assertion Oracles} verify the correctness of the return behavior, although they fail if any exception occurs, which have several common assertion patterns, such as Boolean Assertions (e.g., \texttt{assertTure}, \texttt{assertFalse}), Nullness Assertion (e.g., \texttt{assertNull}, \texttt{assertNotNull}) and Equality Assertions (e.g., \texttt{assertEquals}, \texttt{assertArrayEquals}). 
The \textbf{Tests with Expected Exception Oracles} verify whether the execution of the test prefix with invalid usage can raise an exception. 
Usually, they are frequently expressed with the \texttt{try\{$\cdots$\}} \texttt{catch(Exception e)\{$\cdots$\}} structure or \texttt{assertThrows} statement.
The unit test illustrated in Fig.~\ref{fig:motivation}(b) verifies whether the element is successfully inserted into the target array, while the unit test illustrated in Fig.~\ref{fig:motivation}(c) checks to catch the expected exceptions. 
Based on the example above, we have the following observations:


\textbf{Observation 1.} A unit test is well structured with two parts: a prefix and an oracle. A complete unit test must contain both a test prefix and an oracle. 
Prior works~\cite{sun2023revisiting,zhang2023saga,dinella2022toga} are impractical due to the lack of ability to generate full unit test cases that can be compiled and executed. We applied some SOTAs on the above example and they failed to generate complete tests, for example, EditAS~\cite{sun2023revisiting}. Even though the SOTAs are designed to only generate assertions, such as TOGA~\cite{dinella2022toga}, they are ineffective in generating the correct assertions due to the missing analysis of test prefixes.

%

\textbf{Observation 2.} Test prefixes and oracles should be semantically connected and correctly aligned in the test case generation. 
More precisely, a prefix for the test with an assertion oracle should be paired with a test oracle for the test with an assertion oracle, not the expected exception oracle unit test, and vice versa. 
Prior works~\cite{xie2023chatunitest,tufano2020unit,alagarsamy2023a3test,yuan2023no} are designed to provide end-to-end capabilities of generating complete unit test cases. However, the alignment and different semantics of prefixes and oracles are not well considered and analyzed in the existing approaches, which leads to the incorrect unit test case generation that cannot be complied with and executed. 
For example, AthenaTest~\cite{tufano2020unit} can generate a complete unit test for the above example in Fig.~\ref{fig:motivation}(a), but the prefix and oracle are not matched, thus the generated unit test cannot execute successfully.



\textbf{Observation 3.} The method under test can be complex, in order to generate correct unit test cases, it requires the learning model to have extraordinary learning and understanding capabilities. 
Some recent LLM-based approaches, such as ChatTester~\cite{yuan2023no} and ChatUniTest~\cite{xie2023chatunitest}, leverage the eminent powerful learning capability of ChatGPT in the unit test case generation and can generate the correct test cases for the above example. 
However, the design of their approaches is heavily dependent on ChatGPT's learning capability. 
For example, the prompts designed by ChatUniTest~\cite{xie2023chatunitest} can be as long as 2700 tokens, which is impractical for models with a smaller text window (e.g., 1024 tokens). 
Moreover, even if some models can handle sequences of such length, the lack of strong command understanding capability, such as Code-Llama~\cite{2023codellama}, Pangu-Coder2~\cite{shen2023pangu}, WizardCoder~\cite{luo2023wizardcoder} and StarCoder~\cite{li2023starcoder} makes it difficult for them to generate ideal responses.

Based on the observations above, we have the following insights:

(1) \textbf{Practical Unit Test Generation.} 
Some previous works~\cite{tufano2020unit,alagarsamy2023a3test,dinella2022toga} only focus on generating a part of a unit test, mostly test oracle generation, which is impractical in the real industry setting as it still requires developers/testers to write a large chunk of a unit test case. 
Thus, we aim to provide an end-to-end unit test case generation capability that takes the focal method under test as input and generate a fully executable unit test case with prefix and test oracles.


(2) \textbf{Cascaded Self-directed for Unit Test Generation.} 
Automatically generating test prefixes and test oracle is not a trivial task since it requires understanding the semantics of the method under test. 
Even with powerful understanding capabilities, LLMs still require some guidance in orchestrating target test prefixes and test oracles.
Instead of directly generating test prefixes or test oracles,  several examples can help LLM to better under the faced task, but the quality of examples will highly affect the capabilities. 
Thus, we design an automatic approach that selects effective examples for in-context learning and composes a prompt for unit test generation. 


(3) \textbf{Model-agnostic Framework for Unit Test Generation.}
Most LLM-based unit test generation approaches (e.g., ChatTester~\cite{yuan2023no} and ChatUniTest~\cite{xie2023chatunitest}) highly rely on the powerful learning capability of the close-source model (i.e., ChatGPT), which means that the users have to transfer their sensitive data to that model and then obtain the corresponding responses.
However, privacy data protection is extremely important for individual users, especially for organizations, which prevents the large usage of that type of model.
Thus, we design a model-agnostic framework for unit test generation that can utilize both closed-source models and open-source models, which can ensure better performance and data safety simultaneously.

\label{sec:motivation}

\section{APPROACH}

\begin{figure*}
\centering  
\includegraphics[width=.9\linewidth]{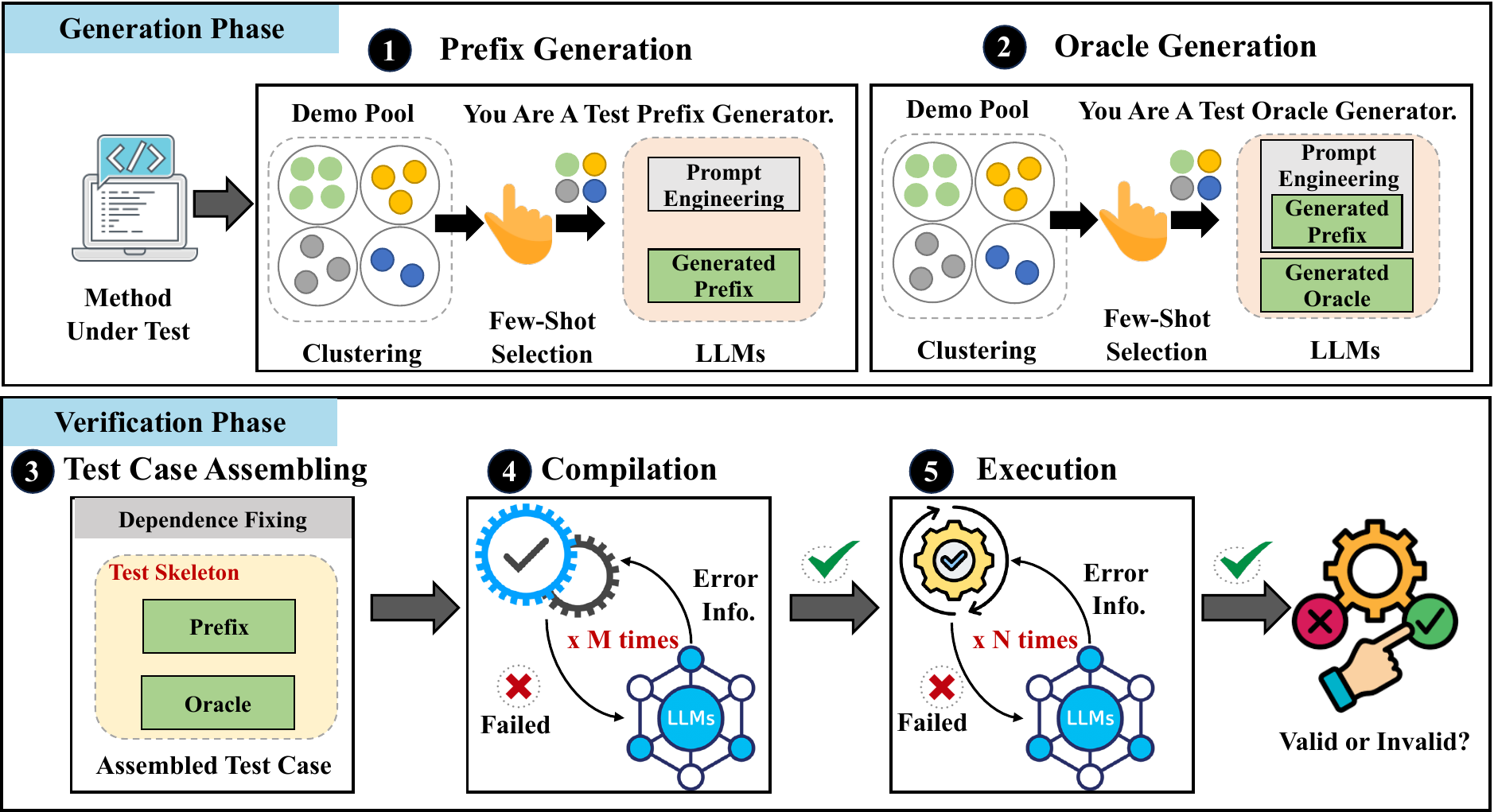}
\caption{The framework of \toolname}
\label{fig:framework}
\end{figure*}

We propose a cascaded self-directed framework, \toolname, for unit test generation. 
\toolname uses instruction tuning combined with few-shot learning to enhance the analysis to understand the semantics of both functions and components in unit tests.
{\bf \toolname} has two main phases (illustrated in Fig.~\ref{fig:framework}): \ding{182} \textbf{generation phase} and \ding{183} \textbf{verification phase}.
The generation phase is to generate the prefix and oracle of a test case. The verification phase verifies the generated test case optionally appending with several attempts to fix compile or execution errors. 
The details of each phase are presented in the following subsections.


\subsection{Generation Phase}

This phase aims to generate separate parts (i.e., prefix and oracle) of a unit test.
To achieve this, we need to address three tasks: (1) Demonstration Pool Construction, (2) Prompt Preparation, and (3) Prefix and Oracle Generation.

\subsubsection{Task 1: Demonstration Pool Construction}
\label{sec:demo_pool_construct}

In few-shot learning~\cite{brown2020language,beltagy2022zero}, high-quality demonstrations are extremely important to help LLMs better understand downstream tasks.
Meanwhile, considering the inherent structure of the method under test written by object-oriented programming language, we split a unit test case into several parts for better representation.
For example,  one Java method is part of a particular class and its corresponding unit test case (i.e., another Java method) usually involves two parts: test prefix and test oracle.
Thus, we separately build two types of demonstration pools:
test prefix demo pool and test oracle demo pool.


In the prefix demo pool, each demo is split into five parts: (1) the class where the focal method is located; (2) the constructor parameters of the class; (3) the signature of the focal method; (4) the name of the test method; (5) the corresponding test prefix written by developers.

\begin{sloppypar}
In the oracle demo pool, each demo can be also split into three parts: (1) the signature of the focal method; (2) the implementation body method under test but WITHOUT the test oracle written by developers (i.e., replaced by a special placeholder of \texttt{<OraclePlaceHolder>}); (3) the corresponding test oracle written by developers.
The structure of the prefix and oracle demo is illustrated in the third part in Fig.~\ref{fig:generation_template} (marked as \ding{174} Few-Shot Selection).
\end{sloppypar}

\subsubsection{Task 2: Prompt Preparation}

\begin{figure*}
\centering  
\includegraphics[width=.85\linewidth]{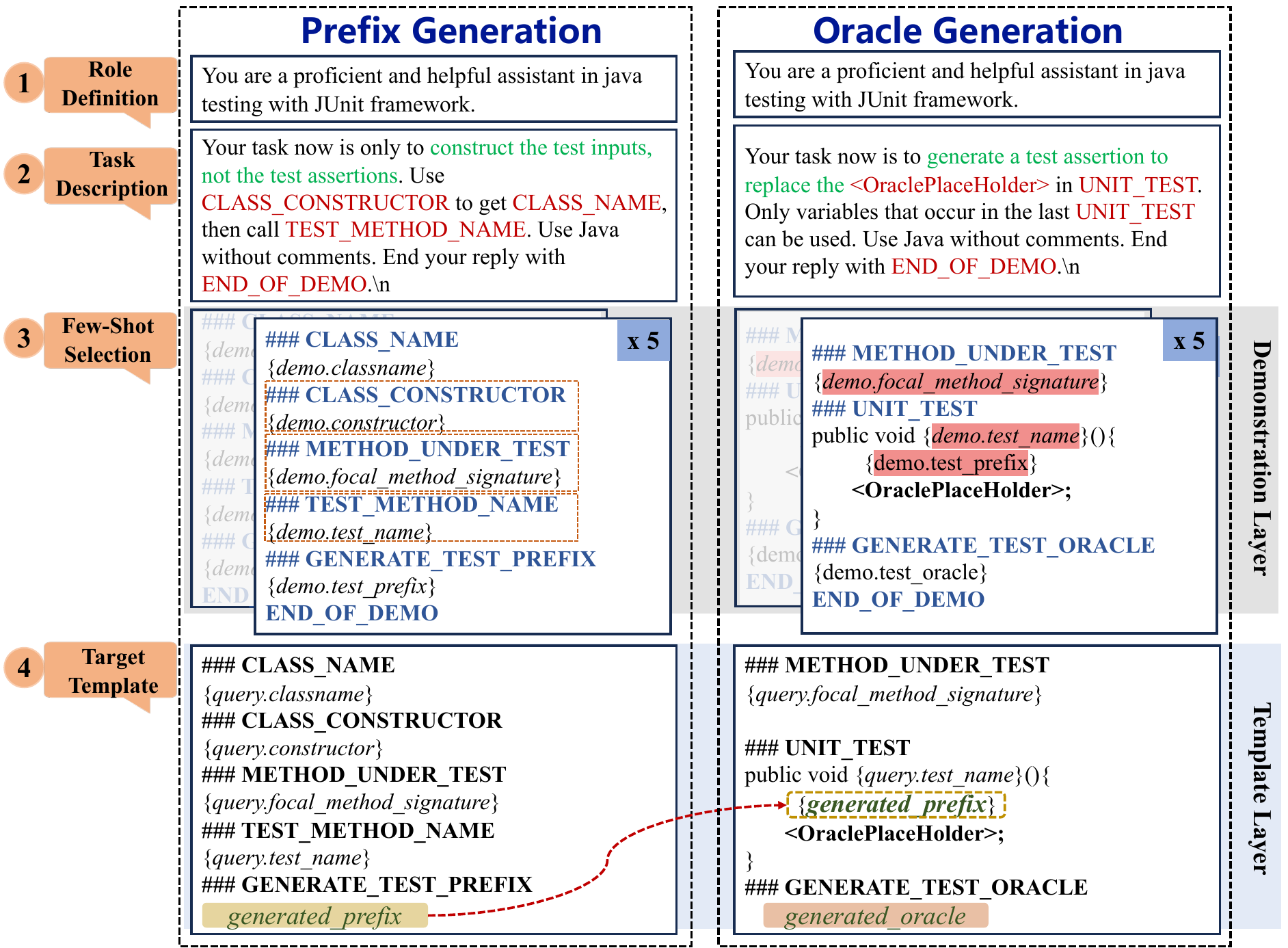}
\caption{The design of generation prompts: involving prefix generation and oracle generation.}
\label{fig:generation_template}
\end{figure*}

The prompt used in \toolname for both test prefix generation and test oracle generation involves four important components as illustrated in Fig.~\ref{fig:generation_template}:

\begin{itemize}[leftmargin=*]
\item \textbf{Role Definition} (marked as \ding{172}). 
\toolname starts a role for a large language model with an instruction like \textit{"You are a proficient and helpful assistant in java testing with JUnit framework"}.
We adopt the same instruction for prefix and oracle generation.

\item \textbf{Task Description} (marked as \ding{173}). 
The LLM is provided with different descriptions for different tasks.
As for prefix generation, we instruct LLMs with such a description like \textit{``Your task now is only to construct the test inputs, not the test assertions. Use \textit{CLASS\_CONSTRUCTOR} to get \textit{CLASS\_NAME}, then call \textit{TEST\_METHOD\_NAME}. Use Java without comments. End your reply with \textit{END\_OF\_DEMO}."}
As for oracle generation, we instruct LLMs with another description like \textit{``Your task now is to generate a test assertion to replace the \textit{<OraclePlaceHolder>} in \textit{UNIT\_TEST}. Only variables that occur in the last \textit{UNIT\_TEST} can be used. Use Java without comments. End your reply with \textit{END\_OF\_DEMO}."}.
Notice that, there are several variables (e.g., \textit{TEST\_METHOD\_NAME}) in both prompts.
They have the same meanings as defined in Section~\ref{sec:demo_pool_construct}.
Meanwhile, the variables with the same names in the prefix prompt and oracle prompt indicate the same instances for building a better connection between the two parts in one test case.

\item \textbf{Few-Shot Selection} (marked as \ding{174}).
LLMs usually need a high-quality prompt to instruct itself to finish the downstream tasks and it is the focus of many works~\cite{xia2023keep,white2023prompt,feng2023prompting}. 
Similarly, we design a structured template for representing a high-quality test case example (i.e., human-written ones), especially, 
 we split a test case into several parts (e.g., driven class name, focal method signature, test prefix, etc.).
To better inspire LLMs' capability, we need to identify the most beneficial examples from the demonstration pool. 
Meanwhile, previous work~\cite{zhang2022automatic} also concludes that a few diverse examples may assist LLMs in achieving a better generalization ability.
To achieve this diversity, \toolname clusters these examples inside the demonstration pool on the basis of their semantic similarity to pick out distinct ones~\cite{zhang2022automatic,li2023mot}.
In particular, we adopt different structures to represent a test case for a better abstraction.
In the prefix generation phase, a test case is composed of the class name (i.e., \textit{demo.classname)}, the constructor of the class (i.e., \textit{demo.constructor}), the signature of the method under test (i.e., \textit{demo.focal\_method\_signature}) and the human-written test prefix (i.e., \textit{demo.test\_prefix}).
In the oracle generation phase, a test case is composed of the signature of the method under test (i.e., demo.focal\_method\_signature), the test method containing placeholders (i.e.,\textit{demo.test\_name}, \textit{demo.test\_prefix} and $<$\textit{OraclePlaceHolder}$>$) and the human-written test oracle (i.e., \textit{demo.test\_oracle}).
Then, each example in the demonstration pool is encoded with a pre-trained language model (i.e., UniXcoder~\cite{guo2022unixcoder}).
Furthermore, considering the limitation of LLMs' conversation windows, we cluster all examples in the demonstration pool into five clusters, and one sample with the highest cosine similarity with the target focal model is picked from each cluster (i.e., five shots used in this paper for both prefix and oracle generation). 
Eventually, three distinct sorting strategies are considered for ordering selected examples, and the details are elaborated as follows. 

    \begin{itemize}
    \item \textbf{Randomly Selection} first select the highest cosine similarity with the target focal method from each cluster and then randomly sort them in the prompt design.
    \item \textbf{Ascending Selection} follows the ``Randomly Selection'' strategy to pick five examples and then sort them in ascending by cosine similarly in the prompt design.
    \item \textbf{Descending Selection} follows the ``Randomly Selection'' strategy to pick five examples and then sort them in descending by cosine similarly in the prompt design.
    \end{itemize}

\item \textbf{Target Template} (marked as \ding{175}). 
Previous work~\cite{nashid2023retrieval} concludes that both the examples and target generation having the same structure in the few-shots learning scenario may yield a better performance.
Therefore, we encode the target method under test in the same way as the examples in corresponding demonstration pools (i.e., prefix and oracle).
\end{itemize}

\subsubsection{Task 3: Prefix and Oracle Generation}
\toolname utilizes the selected examples and the target method under test to construct a prompt (i.e., \ding{172} + \ding{173} + \ding{174} + \ding{175} marked in Fig.~\ref{fig:generation_template}).
Then, \toolname uses this prompt to interact with LLMs and help them infer how to generate the corresponding prefix and oracle solution in order.
\toolname first interacts with LLM through a well-designed prefix prompt and obtains the output: \texttt{generated\_prefix}.
Then, \toolname subsequently makes another interaction with LLM through an oracle prompt with previously generated content (i.e., \texttt{generated\_prefix}) and obtain corresponding output: \texttt{generated\_oracle}.
Notice that, to make a better combination between prefix generation and oracle generation, the variables with the same names in the counterpart layer (i.e., Demonstration Layer and Template Layer) inside a prompt have the same meaning and point to identical content.
For example, in the few-shot selection part (i.e., labeled as ``Demonstration Layer'' in Fig.~\ref{fig:generation_template}), the variable \textit{demo.test\_name} refers to the particular and identical name of a test, which makes a connection between prefix and oracle generation.

\begin{figure*}
\centering  
\includegraphics[width=.85\linewidth]{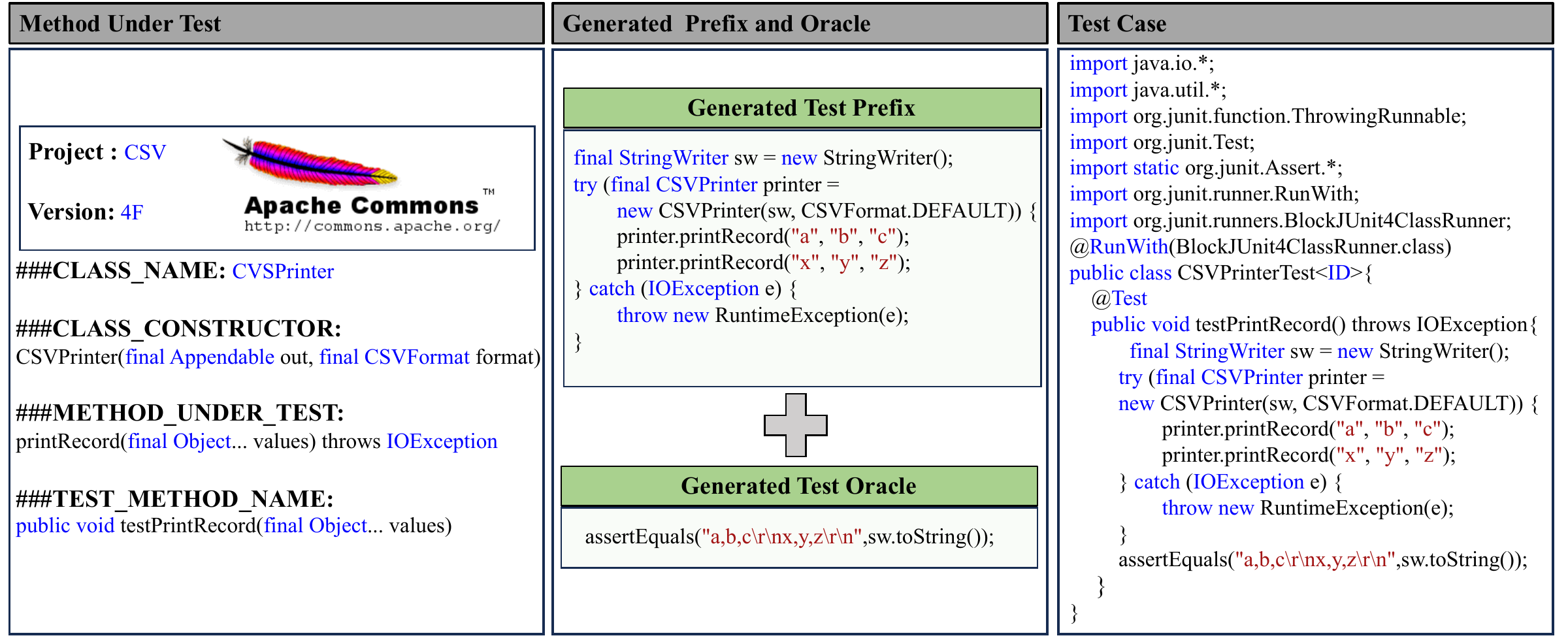}
\caption{The process of test case assembling from test prefix and test oracle}
\label{fig:assembling_process}
\end{figure*}

\subsection{Verification Phase}
This phase aims to verify previously generated separate parts (i.e., prefix and oracle) of a test.
To achieve this, we need to address three tasks: (1) Test Case Assembling, (2) Compilation Verification, and (3) Execution Verification.

\subsubsection{Task 1: Test Case Assembling}
A complete test case is the combination of test prefix and test oracle in an appropriate way optionally appending with dependencies fixing.
In previous steps, the generated prefix and generated oracle are couples of Java statements even with only one statement.
Therefore, we combine the prefix and oracle together and put them into a test method skeleton with the import statements of JUnit framework.
Meanwhile, methods under test may be connected with other classes or libraries.
It is necessary to address the corresponding dependencies with appropriate importing operations (e.g., \texttt{import java.io.File}) or adding annotations (e.g., \texttt{@Test}).
The process of the whole assembling is illustrated in Fig.~\ref{fig:assembling_process} for exampling.
In Fig.~\ref{fig:assembling_process}, there is a method named \texttt{printRecord} under test inside a class named \texttt{CSVPrinter} from Apache Commons project.
This information is organized according to the prompt template shown in the left of Fig.~\ref{fig:assembling_process}.
Then, \toolname interacts with LLMs and obtains two generated parts: test prefix and test oracle.
Both of them are several Java statements shown in the middle of Fig.~\ref{fig:assembling_process}.
Finally, these statements are organized into a testing function (i.e., \texttt{testPrintRecord()} annotated with \texttt{@Test}) nested in testing driven class (i.e., \texttt{CSVPrinterTest}) based on the requirement of JUnit framework and fixing the dependencies by importing a couple of packages.

\subsubsection{Task 2: Compilation Verification}

\toolname compiles the candidate test case to verify the correctness.
If the compilation is successfully executed (e.g., without syntax errors), \toolname will turn to the execution verification stage.
Otherwise, \toolname collects failing test information, which can aid LLMs in understanding the failure causes and provide guidance to fix compilation errors.
Inspired by mutation-based testing~\cite{jain2023contextual,moradi2023effective,zhang2023shapfuzz,louloudakis2023mutatenn}, the generated material may be a good starting point when mutating a good test case even if it may contain a few syntax errors.
Thus, we do not stop at the stage when complication meets error but instead interact with LLMs for a fine-tuned one for better efficiency. 
Then, \toolname reconstructs the prompt, appends the failing information (i.e., illustrated in Fig.~\ref{fig:feedback_complilation_execution}(a)), and feeds it back to the LLM for test case repair. 
The feedback prompt template involves several aspects: (1) the tested method signature, (2) the class containing the focal method, (3) the compiled failure errors, and (4) the originally assembled non-compilable test case (i.e., no interaction with LLM) or previously generated non-compilable test case (i.e., after interacting with LLMs).
Following that, \toolname interacts with LLMs using the new prompt to generate a new candidate test case.
Through feedback error information, LLMs can avoid generating similar mistakes and also learn from previous interactions based on the new prompt.
This iterative process continues until a compilable test case is achieved (i.e., successfully compiled by Java) or exceeds the maximum $M$ number of interactions (i.e., four times for a better balance between effectiveness and cost).

\subsubsection{Task 3: Execution Verification}

Similar to compilation verification, \toolname executes the compilable candidate test case to verify functionality correctness.

\begin{figure*}
\centering  
\includegraphics[width=.85\linewidth]{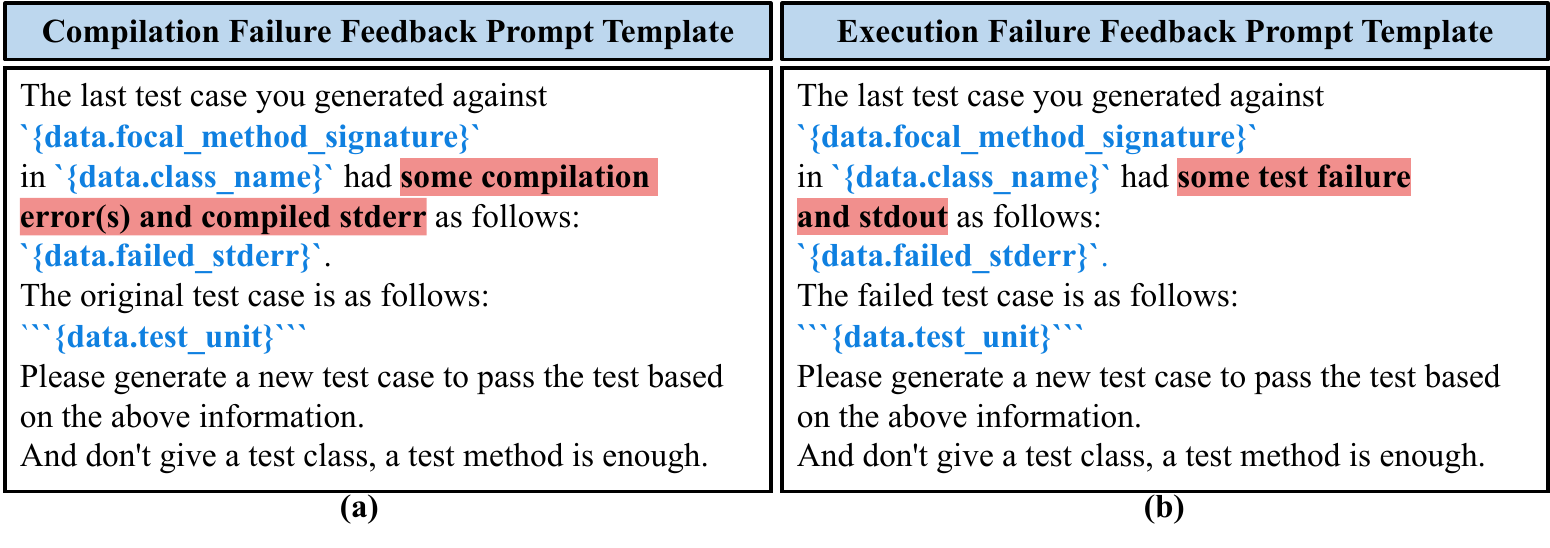}
\caption{The design of feedback prompts during fixing compilation and execution errors with LLMs}
\label{fig:feedback_complilation_execution}
\end{figure*}

If the execution is successful (i.e., free of runtime errors and producing the expected results), we obtain a good test case and the process is terminated.
Otherwise, \toolname also collects failing test information and reconstructs the prompt, appends the failing information (i.e., illustrated in Fig.~\ref{fig:feedback_complilation_execution}(b)), and feeds it back to the LLM for test case repair. 
The feedback prompt template also involves several aspects: (1) the tested method signature, (2) the class containing the focal method, (3) the executed failure errors, and (4) the originally assembled non-executable test case (i.e., no interaction with LLM) or previously generated non-executable test case (i.e., after interacting with LLMs).
This iterative process continues until an executable test case is achieved (i.e., successfully executed by Java) or the maximum $N$ number of interactions exceeds (i.e., three times is used for this study since the context window limitation of LLMs).

\label{sec:approach}


\section{Experimental Setup}
\label{sec:setup}

\subsection{Dataset}

\subsubsection{Studied Dataset}


We follow prior studies ~\cite{tufano2020unit}~\cite{alagarsamy2023a3test}~\cite{xie2023chatunitest} and use Defects4J~\cite{just2014defects4j} as our studied dataset. 
Specifically, we select five popular and commonly evaluated projects included in the dataset: Apache Common Cli~\cite{2023CLI}, Apache Common Csv~\cite{2023CSV}, Google Gson~\cite{2023gson}, JFreeChart~\cite{2023Chart} and Apache Commons Lang~\cite{2023Lang}. These projects represent different domains, namely cmd-line interface, data processing, serialization, visualization and utility. 
Table~\ref{tab:datasets} presents detailed information of our studied projects in the Defects4J dataset.

\begin{table}[htbp]
  \centering
  \caption{Defects4J projects on evaluation}
  \resizebox{\linewidth}{!}
  {
    \begin{tabular}{lllrrr}
    \toprule 
    \textbf{Project}  & \textbf{Abbr.}& \textbf{Domain}& \textbf{\# \makecell[r]{Focal\\Class}}& \textbf{\# \makecell[r]{Focal Method\\(Original)}} & \cellcolor{lightgray} \textbf{\# \makecell[r]{Focal Method\\(Manually Built)}}\\
    \midrule 
    Commons-Cli & Cli & Cmd-line interface  & 13    & 645  & \cellcolor{lightgray}488\\
    Commons-Csv & Csv & Data processing   & 5     & 373 & \cellcolor{lightgray}430\\
    Gson & Gson & Serialization & 15    & 224 &\cellcolor{lightgray}224\\
    Jfreechart & Chart& Visualization & 24    & 1,318 & \cellcolor{lightgray}2,030\\
    Commons-Lang & Lang  & Utility& 28    & 2,712 & \cellcolor{lightgray}4,535\\
    \midrule
    \textbf{All} & &   & \textbf{85}    & \textbf{5,278} &\cellcolor{lightgray} \textbf{7,707}\\
    \bottomrule 
    \end{tabular}%
    }
  \label{tab:datasets}%
\end{table}%


\subsubsection{Demonstration Pool}

To better serve \toolname which is a two-stage approach for unit test generation, we manually separately build two types of demonstration pools.
As introduced in Section~\ref{sec:demo_pool_construct}, we represent one demonstration with several parts by considering the inherent structure among method, class as well as the structure of the unit test case. 

To ensure quality, the demo pools are also built from the unit test cases from the studied datasets, which were written by developers originally.
We first download the source code of each project from GitHub with the latest version for full access.
Following that,  for each test class (i.e., located in the directory of ``src/test''), we retrieve the corresponding test class under test according to the naming patterns based on a deeper analysis.
For example, the test class for the class ``JsonReader'' is most likely named  ``JsonReaderTest''.
Then, we can identify the source code of the method that is under test according to the name of each test case in the test class.
Following that, the signature of the test class's constructor and the signature of the method under test can be obtained.
Finally, we separate the unit test to obtain the corresponding test prefix and test oracle and build the corresponding demonstrations. 

Notably, when dividing each unit test into two parts (i.e., test prefix and test oracle), we encounter two scenarios: one merely contains a test oracle and the other contains more than one test oracle, especially the mixed types of oracles (i.e., assertion oracle and expected exception oracle).
As for the former, we directly extract the test prefix and test oracle and replace the test oracle with a placeholder.
As for the latter, taking the mixed scenario as an example, we need to extract the two types of test oracles individually.
For the part of the test prefixs, we extract them and optionally merge the test prefix located in \texttt{try} statement with the one located in front of the unit test.
Therefore, we obtain two test instances of this scenario and also replace the test oracle with a placeholder.

Meanwhile, to avoid data leaks, we exclude demos that have both the same class name and the method signature when building the pools since these demons may be opted as query examples for the few-shots.
Eventually, we obtain 85 focal classes and 7,707 focal methods as shown in the last column of Table~\ref{tab:datasets}.
Notice that, such a demonstration pool only needs to be built once and can be applicable to other scenarios with the same programming language.

\subsection{Baselines}

To investigate the effectiveness of \toolname, we consider {four} baselines for a comprehensive comparison. We did not compare \toolname with ChatTester~\cite{yuan2023no} due to the unavailability of its source code and the limited description provided in their paper, making it difficult for us to replicate their work.

\begin{itemize}[leftmargin=*]
    \item \textbf{AthenaTest}~\cite{tufano2020unit}. It is the first approach to formulate unit test cases as a sequence-to-sequence learning task. 
    Compared to previous methods, AthenaTest's main contribution is to take the understandability of test cases into consideration rather than solely focusing on code coverage.
    More precisely, it utilizes BART~\cite{lewis2019bart} to denoise pre-training on a Java corpus and then performs supervised fine-tuning on its own dataset (i.e., methods2test~\cite{2021methods2test}). 
    \item \textbf{A3Test}~\cite{alagarsamy2023a3test}. Similar to AthenaTest, A3Test also treats test generation as a sequence-to-sequence task. It fine-tunes PLBART~\cite{ahmad2021unified} specifically for the assertion generation task and incorporates validation mechanisms for naming conventions.
    \item \textbf{ChatUniTest}~\cite{xie2023chatunitest}. It is the first automatic unit test generation method based on  ChatGPT by leveraging the capabilities of large language models to follow natural language instructions. 
    It can perform both verification and repair of test results after test generation. 
    \item \textbf{CAT-LM}~\cite{rao2023cat}. It's a GPT-style aligned code and tests language model with 2.7 billion parameters, trained on a corpus of Python and Java projects. Its key feature is a novel pretraining signal that considers code-test mappings when available. 
\end{itemize}




\subsection{Evaluation Metrics}
We adopt two widely used metrics to evaluate the performance of \toolname and the baseline approaches.
\begin{itemize}[leftmargin=*]
    \item \textbf{Accuracy}. It represents the percentage of correct tests in all generated tests.
    A test case is regarded as correct if and only if it passes and invokes the method under test directly or indirectly.

    \item \textbf{Focal Method Coverage}. 
    It represents the percentage of focal methods that are correctly tested by the generated unit tests. 
    If there is at least one correct attempt in all the generated test cases that tests the focal method correctly, then the method is considered covered.
\end{itemize}

\subsection{Implementation Details}

We implemented \toolname in Python by wrapping the large language models'  ability through the API support (e.g., ChatGPT~\cite{2023chatgptendpoint}) or source code availability (e.g., CodeLlama~\cite{roziere2023code}) and adhere to the best-practice guide~\cite{shieh2023best} for each prompt.
We utilize the \textit{gpt-3.5-turbo} model from the ChatGPT family by default, which is the version used uniformly for our experiments for better generation efficiency. 
Besides, ChatGPT has a maximum input limit of 4,096 tokens, we employ five few-shot examples for prefix generation and oracle generation and set the maximum interaction number for compilation failures to 3 and for runtime failures to 2 during validation.
We also consider several other types of LLMs (e.g., GPT-4) for demonstrating the generalization of our model, especially the open-source ones: DeepSeek-Coder~\cite{2023deepseek-coder}, CodeLlama~\cite{roziere2023code} and its variant~\cite{2023phindcodellama}.

{
For baselines, the experimental results on the prior study of Xie et al.~\cite{xie2023chatunitest} are directly adopted for the first three methods due to consistent benchmarks. 
For CAT-LM, a replication is conducted.
The replication process involves initially locating the model~\cite{nikitharao2023catlm} and data processing code~\cite{Zenodohome2023catlm}. 
By combining the content of the paper with the outcomes of executing the data processing code, the appropriate prompts to be provided to CAT-LM are determined. Considering CAT-LM's approach to generate code at the granularity of entire classes for generating test classes, whereas Defects4j requires the generation of test cases for individual methods, other public methods in the class under test are removed from the code repository for each query. 
For the evaluation of results, a method is employed where test cases are generated ten times for each query, all test cases are aggregated, and the generation is considered successful if at least one correct test case exists among them.}



\label{sec:experiemt}

\section{Results}
\label{lab:results}
 
We present the results by proposing and answering the following three research questions:

\begin{itemize}[leftmargin=*]

\item \textbf{RQ1 Effectiveness Comparison}. { {How is the performance of \toolname compared to state-of-the-art (SOTA) approaches? }}

\item \textbf{RQ2 Cascade and Model-agnostic}. { {How is the performance of \toolname when generating in a cascaded manner compared to non-staged generation using the same model?
Does the efficacy of \toolname hinge on particular large language models?}}

\item \textbf{RQ3 Order of Demonstrations}. { {How does the order of demonstrations inside a prompt affect the performance of \toolname?}}


\end{itemize}


\subsection{RQ1:  Effectiveness Comparison}
\label{lab:rq1}


\begin{table}[htbp]
  \centering
  \caption{Accuracy of \toolname compared with the baselines (RQ1).}
    \resizebox{1.00\linewidth}{!}
  {
    \begin{tabular}{lrrrrr}
\toprule
   \textbf{Projects} & \textbf{AthenaTest} & \textbf{A3Test} & \textbf{ChatUniTest} & \textbf{CAT-LM} & \textbf{\toolname$_{GPT3.5}$} \\
   \midrule
   Cli   & 11.07\% & 25.19\% & 38.77\% & 45.60\% & \textbf{58.29\%} \\
   Csv   & 8.98\%  & 25.73\% & 44.26\% & 43.66\% & \textbf{69.17\%} \\
   Gson  & 2.89\%  & 14.09\% & 23.3\% & 24.17\% & \textbf{63.18\%} \\
   Chart & 11.7\%  & 31.3\%  & 39.02\% & 37.84\% & \textbf{76.66\%} \\
   Lang  & 23.35\% & 49.5\%  & 39.85\% & 56.15\% & \textbf{84.13\%} \\
\midrule
   \rowcolor{lightgray}\textbf{Total} & {16.21\%} & {40.05\%} & {40.41\%} & {48.04\%} & \textbf{77.16\%} \\
   \bottomrule
   \end{tabular}%
   }
  \label{tab:rq1_accuracy}%
\end{table}%

\phead{Motivation.}
In this RQ, we aim to investigate the overall performance of \toolname on unit test generation by comparing the results with four baseline approaches discussed in Section~\ref{sec:setup}.

\phead{Approach.}
We use the datasets discussed in Section~\ref{sec:setup} to conduct the experiments using \toolname and compare the accuracy with the baselines. For each focal method, we extract its corresponding class name, constructor signature, and method signature from the datasets to generate unit tests with \toolname. 
We use \textit{gpt-3.5-turbo} as the backbone model of \toolname.



\begin{table}[htbp]
  \centering
  \caption{Focal method coverage of \toolname compared with the baselines (RQ1).}
      \resizebox{1.00\linewidth}{!}
  {
    \begin{tabular}{lrrrrr}
    \toprule
    \textbf{Projects} & \textbf{AthenaTest} & \textbf{A3Test} & \textbf{ChatUniTest} & \textbf{CAT-LM} & \textbf{\toolname$_{GPT3.5}$} \\
    \midrule
    Cli   & 29.46\% & 37.2\% & \textbf{70.34\%} & 59.22\% & 61.98\% \\
    Csv   & 34.31\%  & 37.8\% & 76.94\% & 66.76\% & \textbf{77.42\%} \\
    Gson  & 9.54\%  & 40.9\% & 55.91\% & 47.27\% & \textbf{63.35\%} \\
    Chart & 32.00\%  & 34.40\%  & 79.56\% & 49.25\% & \textbf{82.55\%} \\
    Lang  & 56.97\% & 58.30\%  & 84.05\% & 70.06\% & \textbf{88.49\%} \\
    \midrule
    \rowcolor{lightgray}\textbf{Total} & 43.75\% & 46.80\% & 79.67\% & 62.32\% & \textbf{81.93\%} \\
    \bottomrule
    \end{tabular}%
    } 
  \label{tab:rq1_focal_method_coverage}%
\end{table}%

\phead{Results.} Overall, \textbf{we find that \toolname outperforms the baseline approaches in terms of both accuracy and focal method coverage.} Below, we discuss the results of accuracy and focal method coverage, respectively.

\noindent \underline{Accuracy.}
Table~\ref{tab:rq1_accuracy} shows the percentage of test cases that are correctly generated on five projects (i.e., accuracy). The best result among different approaches is marked in bold. 
Overall, \toolname achieves the accuracy (i.e., 77.16\% on average), while the average accuracy of the baselines ranges from 16.21\% for {\sf AthenaTest} to 48.04\% for {\sf CAT-LM}.
Upon examining some of the test methods successfully generated by \toolname, which other baselines failed to accomplish, it is found that generating the test prefix independently allows the LLM to better handle the instantiation of classes under different design patterns.

\noindent \underline{Focal Method Coverage.}
Table~\ref{tab:rq1_focal_method_coverage} presents the results of focal method coverage for \toolname and the three baselines.
Notably, the performance of \toolname is obtained with a single round with LLM, which means \toolname's pipeline is executed only once for each item in the dataset. 
In contrast, ChatUniTest's performance is a cumulative result of six rounds, and the other two baselines' results are also the accumulation of several dozen rounds.
Overall, \toolname outperforms all the baselines on four out of the five studied projects (i.e., {\sf Csv}, {\sf Gson}, {\sf Chart} and {\sf Lang}), with an average focal method coverage of 81.93\%. The average focal method coverage of the three baselines ranges from 43.75\% for {\sf AthenaTest} to 79.67\% for ChatUniTest.

Although accuracy and focal method coverage generally show a positive correlation, some results deviate from this pattern. Despite CAT-LM performing ten generations for each focal method coverage, the final set of all correct test cases does not stand out in terms of focal method coverage. It is speculated that this is because CAT-LM, as a language model with a smaller parameter size that has been trained on five test projects, tends to overfit, which is not conducive to generating more diversified test cases.

As for \texttt{Cli} project, ChatUniTest performs better than \toolname.
Through deep analysis, we find that  
ChatUniTest generates a test suite (usually containing several unit tests) for one focal method,  while \toolname just generates one unit test for a particular focal method.
As an example shown in Fig.~\ref{fig:good_case}, \toolname generates only unit test for the target method but ChatUniTest generates two.
Therefore, it brings a higher possibility of covering the focal method to ChatUniTest.

\intuition{{\bf Summary of RQ1: } For the effectiveness of unit test generation, \toolname achieves an accuracy of 77.16\% and a focal method coverage of 81.93\% on average, which outperforms the three baselines. }

\subsection{RQ2: Cascade and Model-agnostic}
\label{lab:rq2}

\begin{table}[htbp]
 
  \centering
  \caption{Results of cascaded and direct pipelines using \toolname$_{GPT3.5}$ (RQ2)}
  \resizebox{\linewidth}{!}
  {
    \begin{tabular}{lr|rr|rr}
    \toprule
    \multirow{2}[0]{*}{\textbf{Projects}}     & \multirow{2}[0]{*}{\textbf{\# Query}}       & \multicolumn{2}{c|}{\textbf{\% Accuracy}} & \multicolumn{2}{c}{\textbf{\% Focal Method Coverage}} \\
    \cmidrule{3-6}
     &  & {\textbf{Cascaded}} & {\textbf{Direct}} & {\textbf{Cascaded}} & {\textbf{Direct}} \\
    \midrule
    Cli   & 645   & \textbf{58.29\%} & 51.32\% & \textbf{61.98\%} & 53.02\% \\
    Csv   & 373   & \textbf{69.17\%} & 61.93\% & \textbf{77.42\%} & 69.44\% \\
    Gson  & 220   & \textbf{63.18\%} & 50.91\% & \textbf{63.35\%} & 53.18\% \\
    Chart & 1,328  & \textbf{76.66\%} & 69.20\% & \textbf{82.55\%} & 75.15\% \\
    Lang  & 2,712  & \textbf{84.13\%} & 73.64\% & \textbf{88.49\%} & 78.47\% \\
    \midrule
    \textbf{All}   & 5,278  & \textbf{77.16\%} & 68.02\% & \textbf{81.93\%} & 72.83\% \\
    \bottomrule
    \end{tabular}%
    }
  \label{tab:rq2_cascaded_and_nonstop_chatgpt3.5}%
\end{table}%

\begin{table*}[htbp]
  \centering
  \caption{Comparison of accuracy between different models (RQ2). }
  \resizebox{1\linewidth}{!}{
    \begin{tabular}{lr|rr|rrr|rrrr}
    \toprule
    \multicolumn{1}{c}{\multirow{2}[1]{*}{\bf Projects}} & \multicolumn{1}{c|}{\multirow{2}[1]{*}{\bf \# Query}} & \multicolumn{8}{c}{\bf \% Accuracy} \\
    \cmidrule{3-11}      &       & \multicolumn{1}{l}{\toolname$_{GPT4.0}$} & \multicolumn{1}{l|}{\toolname$_{GPT3.5}$} & \multicolumn{1}{l}{\toolname$_{ds}$} & \multicolumn{1}{l}{\toolname$_{phind-cl}$} & \multicolumn{1}{l|}{\toolname$_{cl}$} & \multicolumn{1}{l}{ChatUniTest} & \multicolumn{1}{l}{A3Test} & \multicolumn{1}{l}{AthenaTest} & \multicolumn{1}{l}{CAT-LM} \\
            \midrule
            Cli  & 645   & \textbf{66.82\%} & \textbf{58.29\%} & 35.00\% & 27.47\% & 16.43\% & 38.77\% & 25.19\% & 11.07\% & 45.60\% \\
            Csv & 373   & \textbf{76.68\%} & \textbf{69.17\%} & \textbf{51.20\%} & \textbf{54.95\%} & 37.00\% & 44.26\% & 25.73\% & 8.98\% & 43.66\% \\
            Gson & 220   & \textbf{95.45\%} & \textbf{63.18\%} & \textbf{45.00\%} & \textbf{30.85\%} & \textbf{24.09\%} & 23.30\% & 14.09\% & 2.89\% & 24.17\% \\
            Chart  & 1,328  & \textbf{87.80\%} & \textbf{76.66\%} & \textbf{72.10\%} & \textbf{44.37\%} & \textbf{40.59\%} & 39.02\% & 31.30\% & 11.70\% & 37.84\% \\
            Lang  & 2,712  & \textbf{89.71\%} & \textbf{84.13\%} & \textbf{76.40\%} & \textbf{60.32\%} & \textbf{51.47\%} & 39.85\% & 49.50\% & 23.35\% & 56.15\% \\
            \midrule
          \rowcolor{lightgray}  Total & 5,278  & \textbf{85.75\%} & \textbf{77.16\%} & \textbf{67.17\%} & \textbf{50.68\%} & \textbf{42.29\%} & 40.41\% & 40.05\% & 16.21\% & 48.04\% \\
    \bottomrule
    \end{tabular}%
    }
  \label{tab:discussion_diff_llm_accuracy}%
\end{table*}%

\phead{Motivation.}
In RQ1, we use \textit{gpt-3.5-turbo} as the backbone model to study the effectiveness of \toolname compared with the baselines.
The results show that \toolname outperforms existing baselines and achieves promising performance for unit test case generation.
In this RQ, we further study the effectiveness of \toolname from two aspects: \textbf{Cascade} and \textbf{Model-agnostic}.
For the aspect of Cascade, we study the effectiveness of \toolname's cascaded pipeline to the general pipeline of directly generating the complete unit test cases. 
For Model-agnostic, we study if \toolname is effective when using different models.


\phead{Approach.} 
We design different experiments to study the aspects of cascade and model-agnostic of \toolname, respectively.

\noindent \underline{Cascade.}
We use \textit{gpt-3.5-turbo} as the backbone model to compare the results of \toolname's cascaded pipeline (i.e., generate test prefix and test oracle in order) with the direct pipeline (i.e., generate the complete unit test case directly). 
Specifically, for the cascaded pipeline, we follow the same setting of RQ1 in using \toolname. 
For the direct pipeline, we generate the complete unit test case directly by replacing the part of the test prefix (i.e., as shown in Figure~\ref{fig:generation_template}) with the complete body of the unit test case.


\noindent \underline{Model-agnostic.}
Apart from \textit{gpt-3.5-turbo}, we use additional state-of-the-art open-source LLMs to investigate if \toolname is still effective when using other models (i.e., model-agnostic). Specifically, the additional models include \textit{gpt-4.0} and three open-source LLMs: (1) \textit{CodeLlama-Instruct}~\cite{2023codellama}, (2) \textit{Phind-CodeLlama-Instruct}~\cite{2023phindcodellama}, and (3) \textit{DeepSeek-Coder-Instruct}~\cite{2023deepseek-coder}. 
We follow the experimental setup discussed in Section~\ref{sec:setup} and compare the results of each model with the baselines. 
When conducting experiments using backbone models, we ensure that the demo selection technique, cascaded strategy, and the number of fixes in the feedback stage remain consistent. 
The results of \toolname in using these three models are referred as \toolname$_{cl}$, \toolname$_{phind-cl}$, and \toolname$_{ds}$, respectively.


\phead{Results.} We discuss the results from the aspects of Cascade and Model-agnostic, respectively.

\noindent \underline{Cascade.}
Table~\ref{tab:rq2_cascaded_and_nonstop_chatgpt3.5} presents the results of unit test generation using \toolname's cascaded pipeline and the direct pipeline. Overall, \textbf{we find \toolname's cascaded pipeline achieves higher accuracy and focal method coverage in all the five studied projects.}
For example, in the {\sf Cli} project, the accuracy of cascaded and direct pipelines are 58.29\% and 51.32\% respectively. The focal method coverage of cascaded and direct pipelines are 61.98\% and 53.02\%, respectively.
In the {\sf Gson} project, we observe the largest difference between the results of cascaded and direct pipelines among the five projects, with the cascaded pipeline outperforming the direct pipeline by 24.10\% in accuracy and 19.12\% in focal method coverage.



\noindent \underline{Model-agnostic.}
Table \ref{tab:discussion_diff_llm_accuracy} and Table \ref{tab:discussion_diff_llm_fmc} present the results of \toolname in using different backbone models in terms of accuracy and focal method coverage. The numbers that are higher than all baselines are marked in bold. Overall, \textbf{we find that \toolname can mostly outperform all the three baselines when using different backbone models.} For accuracy on average, the results of \toolname in using different models range from 42.29\% of \toolname$_{cl}$ to 85.75\% of \toolname$_{GPT4.0}$, which outperform all the three baselines. For focal method coverage on average, the results range from 64.72\% of \toolname$_{cl}$ to 91.85\% of \toolname$_{GPT4.0}$, which outperform two of the three baselines (i.e., except for ChatUniTest).
Among the three open-source models, \textit{DeepSeek-Coder-Instruct} achieves the best overall accuracy (i.e., 67.17\% on average) and \textit{Phind-CodeLlama-Instruct} achieves the best overall focal method coverage (i.e., 87.04\% on average).
\begin{table*}[htbp]
  \centering
  \caption{Comparison of focal method coverage between different models (RQ2).}
  \resizebox{1\linewidth}{!}{
    \begin{tabular}{lr|rr|rrr|rrrr}
    \toprule
    \multicolumn{1}{c}{\multirow{2}[0]{*}{\bf Projects}} & \multicolumn{1}{c|}{\multirow{2}[0]{*}{\bf \# Query}} & \multicolumn{8}{c}{\bf \% Focal Method Coverage} \\
          \cmidrule{3-11}      &       & \multicolumn{1}{l}{\toolname$_{GPT4.0}$} & \multicolumn{1}{l|}{\toolname$_{GPT3.5}$} & \multicolumn{1}{l}{\toolname$_{ds}$} & \multicolumn{1}{l}{\toolname$_{phind-cl}$} & \multicolumn{1}{l|}{\toolname$_{cl}$} & \multicolumn{1}{l}{ChatUniTest} & \multicolumn{1}{l}{A3Test} & \multicolumn{1}{l}{AthenaTest} & \multicolumn{1}{l}{CAT-LM} \\
    \midrule
    Cli   & 645   & \textbf{75.97\%} & 61.98\% & 55.35\% & 58.76\% & 39.38\% & 70.34\% & 37.20\% & 29.46\% & 59.22\% \\
    Csv   & 373   & \textbf{87.40\%} & \textbf{77.42\%} & 75.34\% & \textbf{92.23\%} & 51.74\% & 76.94\% & 37.80\% & 34.31\% & 66.76\% \\
    Gson  & 220   & \textbf{95.45\%} & \textbf{63.35\%} & \textbf{59.55\%} & \textbf{72.27\%} & 40.45\% & 55.91\% & 40.90\% & 9.54\% & 47.27\% \\
    Chart  & 1,328  & \textbf{94.95\%} & \textbf{82.55\%} & \textbf{87.35\%} & \textbf{87.65\%} & 55.65\% & 79.56\% & 34.40\% & 32.00\% & 49.25\% \\
    Lang  & 2,712  & \textbf{94.43\%} & \textbf{88.49\%} & \textbf{93.69\%} & \textbf{93.95\%} & 78.95\% & 84.05\% & 58.30\% & 56.97\% & 70.06\% \\
    \midrule
   \rowcolor{lightgray} Total & 5,278  & \textbf{91.85\%} & \textbf{81.93\%} & \textbf{84.69\%} & \textbf{87.04\%} & 62.32\% & 79.67\% & 46.80\% & 43.75\% & 62.32\% \\
    \bottomrule
    \end{tabular}%
    }
  \label{tab:discussion_diff_llm_fmc}%
\end{table*}%

\intuition{{\bf Summary of RQ2: }\toolname's cascaded pipeline outperforms the direct pipeline by a large margin. Besides, \toolname can also achieve promising results when using additional open-source LLMs, which shows the capability of model-agnostic in unit test generation.}

\subsection{RQ3: Impacts of Demonstration Sorting}
\label{lab:rq3}


\begin{table*}[htbp]
  \centering
  \caption{Comparison between different orders of demos (RQ3).}
  \resizebox{1\textwidth}{!}{
  \begin{threeparttable}
    \begin{tabular}{lr|rrrr|rrrr|rrrr}
    \toprule
     \multirow{2}[1]{*}{\textbf{Projects}}     & \multirow{2}[1]{*}{\textbf{\# Query}}     & \multicolumn{4}{c}{\textbf{\% Accuracy}} & \multicolumn{4}{c}{\textbf{\% Focal Method Coverage}} & \multicolumn{4}{c}{\textbf{\# Average Repair Attempts}} \\
    \cmidrule{3-14}
     &  & {{Random}} & {Ascending} & {{Descending}} & {{Totally Random}} & {{Random}} & {{Ascending}} & {{Descending}} & {{Totally Random}} & {{Random}} & {{Ascending}} & {{Descending}} & {{Totally Random}} \\
    \midrule
    Cli   & 645   & \textbf{58.29\%} & 53.02\% & 55.04\% & 46.05\% & \textbf{61.98\%} & 57.83\% & 61.40\% & 50.23\% & 1.237 & \textbf{1.229} & 1.234 & 1.327 \\
    Csv   & 373   & 69.17\% & \textbf{69.44\%} & 68.63\% & 64.08\% & 77.42\% & \textbf{82.31\%} & 76.94\% & 75.60\% & \textbf{1.011} & 1.091 & 1.172 & 1.196 \\
    Gson  & 220   & \textbf{63.18\%} & 62.73\% & 59.09\% & 59.55\% & 63.35\% & \textbf{63.64\%} & 61.82\% & 60.91\% & \textbf{1.814} & 1.982 & 1.905 & 1.941 \\
    Chart & 1328  & 76.66\% & 76.51\% & \textbf{80.05\%} & 75.15\% & 82.55\% & 83.28\% & \textbf{84.56\%} & 78.61\% & 0.672 & 0.666 & \textbf{0.622} & 0.697 \\
    Lang  & 2712  & \textbf{84.13\%} & 84.07\% & 83.63\% & 79.13\% & 88.49\% & \textbf{88.68\%} & 86.58\% & 83.15\% & 0.501 & \textbf{0.481} & 0.489 & 0.537 \\
    \midrule
    \rowcolor{lightgray} All   & 5278  & \textbf{77.16\%} & 76.45\% & 77.15\% & 72.21\% & 81.93\% & \textbf{82.06\%} & 81.28\% & 76.52\% & 0.725 & 0.725 & \textbf{0.721} & 0.779 \\
    \bottomrule
    \end{tabular}%
     \end{threeparttable}
    }
  \label{tab:rq3_diff_orders}%
\end{table*}%

\phead{Motivation.}
Prior studies investigated the construction of in-context examples in the prompts and observed that the order of examples in the prompts can have a significant impact on the results~\cite{lu2021fantastically}.
In this RQ,  we study the impact of the order of demonstrations in the few-shot examples on \toolname's performance of unit test generation.

\phead{Approach.}
As discussed in Section~\ref{sec:approach}, \toolname retrieves similar examples based on the cosine similarity between the UniXCoder-encoded vectors of the query and demonstrations as the few-shot examples. We cluster all the examples in the demonstration pool into five clusters and select one sample from each of the clusters. 
We use three different strategies to sort the five selected examples based on their cosine similarity when composing the few-shot examples in the prompt: \textit{Random, Ascending, and Descending}. We use \textit{gpt-3.5-turbo} as the backbone model to study the results when using different sorting strategies.
Additionally, we consider another setting where samples are obtained completely randomly, regardless the similarity.



\phead{Results.}
Table~\ref{tab:rq3_diff_orders} shows the results of unit test generation using three sampling orders (Random, Ascending, and Descending) and the completely random extraction way. 
Apart from accuracy and focal method coverage, we also discuss Average Repair Attempts, which are the times of attempts required by the LLM to repair the test cases in the verification stage.

\noindent \underline{Accuracy.}
Across all the studied projects, the Random sampling order demonstrates a small advantage, with three projects (Cli, Gson and Lang) achieving the highest accuracy.
While the Ascending order closely follows the Random order in terms of accuracy, the Descending order exhibits a notable performance in the Chart project, reaching 80.05\%. 
Overall, the Random sorting strategy slightly outperforms the other two with an average accuracy of 77.16\%.
The accuracy obtained with the totally random order is the lowest, being on average 6.12\% lower than the first three arrangements that underwent similarity filtering. This indicates that choosing demos as similar as possible to the query is very important when guiding the LLM in test generation.

\noindent \underline{Focal Method Coverage.} 
Contrary to the results for accuracy, the highest focal method coverage is achieved when demos in the context are sorted in ascending order. It is due to the fact that: when demos closer to the query have higher similarity, the LLM is more likely to invoke methods that share the same name but have different parameters as the method under test, and these methods happen to be included in the benchmark, thereby increasing the focal method coverage. Moreover, the performance obtained with the totally random order remains the worst, as ensuring accuracy is a prerequisite for the coverage of the method under test. 

\noindent \underline{Average Repair Times.}
The differences in repair attempts across various projects are pronounced. For the projects Cli, Csv, and Gson, the average attempts for all four sorting methods exceed one attempt, whereas for the Chart and Lang projects, the average attempts for all sorting methods are below 0.7. This metric indirectly reflects the varying difficulties in generating executable test cases that pass across different projects. The first three sorting strategies, which have undergone similarity filtering, perform similarly, whereas the totally random sorting strategy clearly requires more repair attempts.

\intuition{{\bf Summary of RQ3}: \toolname is not sensitive to the order of cases after clustering and filtering out examples with high similarity. Completely disregarding the similarity between examples, as well as between the examples and the query, and randomly selecting an equal number of examples result in the worst performance across all three metrics. Accuracy and focal method coverage show a positive correlation in the results. Average repair attempts exhibit a negative correlation with the first two metrics.}




\label{sec:results}

\section{Discussion}

\subsection{Case Study}
We conduct a case study based on our experimental results in Section~\ref{lab:results}.
Specifically, we study the correct and incorrect test cases generated by \toolname, respectively.


\subsubsection{Correct Test Case Generated by \toolname}
We exemplify with \texttt{flush()} method in class \texttt{JsonWritter} from \texttt{Gson}~\cite{2023gson} project and its signature is shown as ``\texttt{public void flush() throws IOException}".
We make a unit test generation comparison between \toolname and the current state-of-the-art approach ChatUnitTest.

\begin{figure}  
\centering
\includegraphics[width=\linewidth]{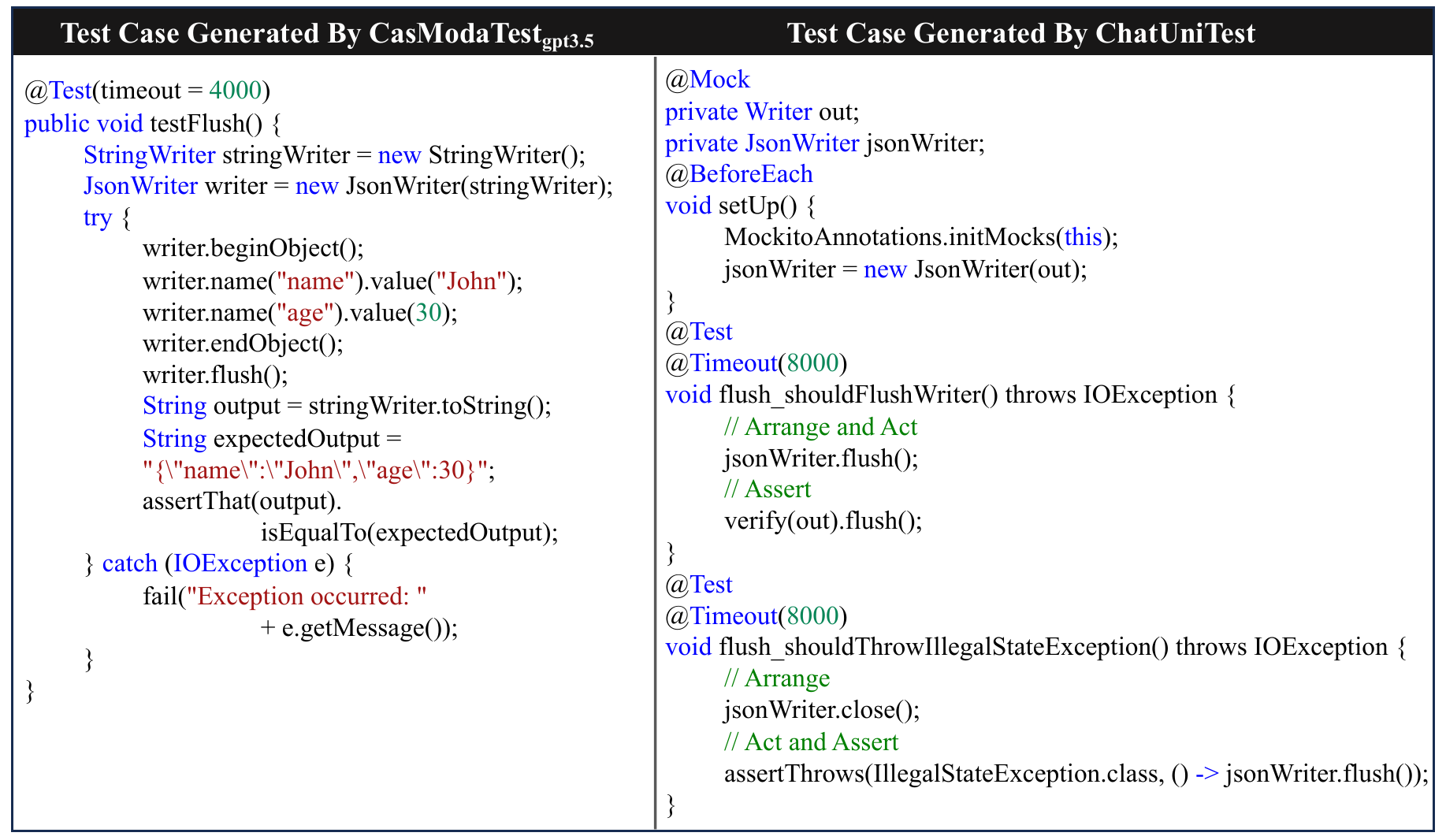}
\caption{The Test Cases Generated by \toolname and ChatUniTest for method flush()}
\label{fig:good_case}
\end{figure}

Fig.~\ref{fig:good_case} shows the two code snippets generated by \toolname and ChatUniTest, respectively and we showcase only the segment within the test method since the space limitation.
Overall, \toolname$_{GPT3.5}$ generates an executable test method after incorporating one fix in the verification phase, while ChatUniTest~\cite{2023chatunitest-maven-plugin} fails to generate any executable test cases after 6 times interaction with ChatGPT by setting \textit{maxRounds} with six according to their paper.

More precisely, \toolname generates an independent test method with well-constructed test prefix and test oracle. 
However, due to the missing of IOException handling, \toolname initially fails to pass the first compilation. 
But this error is easily rectified by the paradigm we defined during the feedback phase and \toolname adds a ``\texttt{try...catch...}" structure to address that error.
Meanwhile, the generated test case has good readability, which helps programmers easily reproduce the bug.

As for ChatUniTest, it utilizes Mockito~\cite{2023mockito} to create a mock object of the output stream and instantiates the ~\texttt{JsonWriter} before each test method is called. 
ChatUniTest totally generates two test methods. 
The first one can successfully execute 
but has limited testing effectiveness since it can only verify whether the \texttt{flush()} method is called. 
However, the second one fails to compile since ChatUniTest misjudged the exception type in the \texttt{assertThrows} method.
It finally fails to complete the repair within a limited number of attempts since the threw exception is mismatched with the one declared in the testing method signature.

\subsubsection{Incorrect Test Case Generated by \toolname}
\begin{figure}[htp]

\centering
\includegraphics[width=\linewidth]{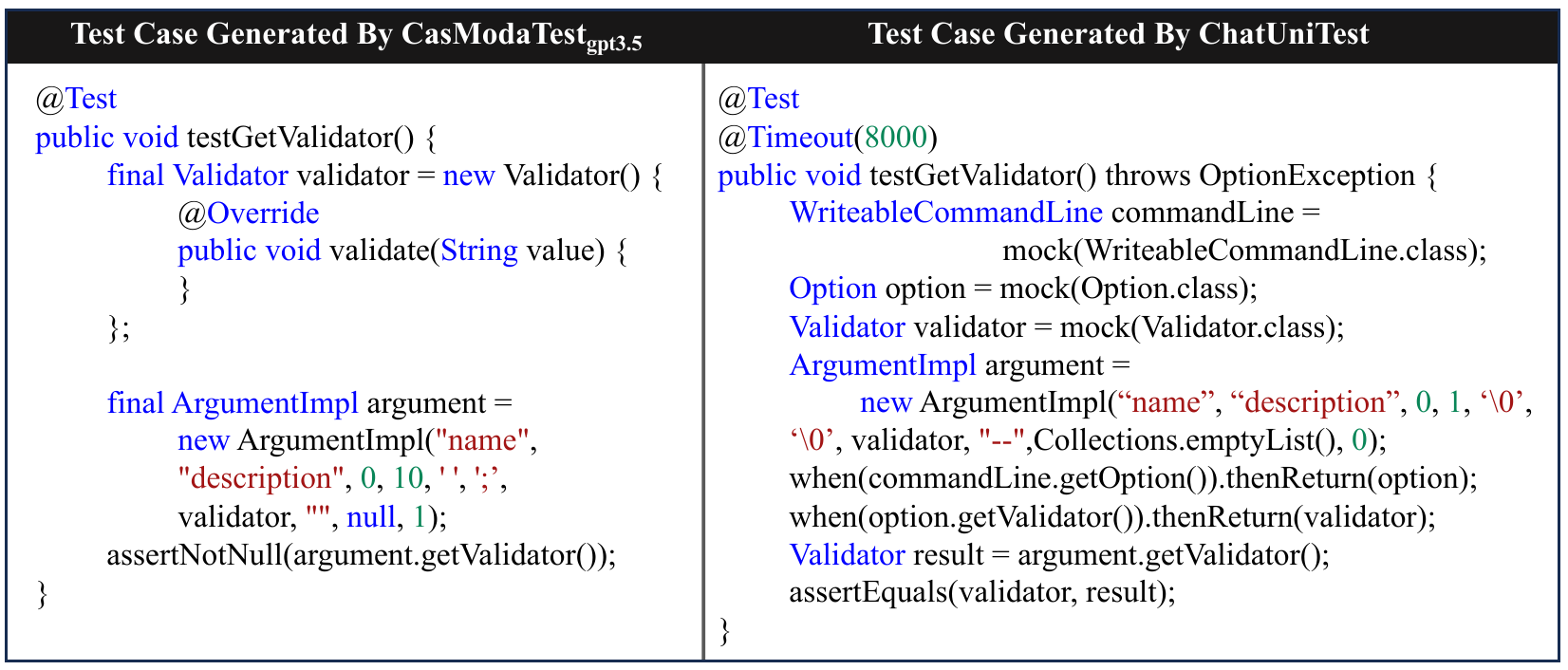}
\caption{The Test Case Generated by \toolname for method getValidator()}
\label{fig:bad_case}

\end{figure}

We exemplify with another method named \texttt{getValidator()} in \texttt{ArgumentImpl} class from \texttt{Cli} project.
However, \toolname fails to generate an executable test case (shown in Fig.~\ref{fig:bad_case}). 
Through deep analysis, we find that the method \texttt{validator} inside an interface should be overridden with the signature of ``\texttt{public void validate(List value)}. 
However, we do not provide \toolname with such information and \toolname cannot be obtained from the compiler's standard error during the subsequent feedback phase. 
Thus, \toolname ultimately fails to generate a test case that passes the compilation within a limited number of attempts (i.e., three fix opportunities).
We also evaluate ChatUniTest on the same method and set the number of interactions with ChatGPT to six. 
ChatUniTest attempts to generate three times, but it fails to generate any that can pass. 
As shown in Fig.~\ref{fig:bad_case}, the latest fixed version of the test method still encounters the issue: "Cannot resolve method 'getValidator' in 'Option'".

\subsection{The efficiency of \toolname}
To explore the efficiency of \toolname, 145 data points from the \texttt{Chart} project are randomly selected for a timing experiment using \textit{gpt-3.5-turbo} as the backbone model. It took a total of 44 minutes and 2 seconds to generate test cases for the selected dataset, resulting in 102 correct test cases. The average time per tested method was 18.2 seconds, and the average time to generate each correct test case was 25.9 seconds.


\subsection{Cross-project Demo Selection}
We conduct a cross-project demo selection study to show the generalization of \toolname.
20\% of the data from both the \texttt{Csv} and \texttt{Gson} projects are extracted, and they swap demo pools for retrieval and generation. Similarly, 10\% of the data from both the \texttt{Lang} and \texttt{Chart} projects are extracted for the same experiment. Table~\ref{tab:cross_pro_selection} shows that  retrieving demos from other projects in the stage of generation may leads to a certain decrease in both accuracy and focal method coverage metrics (drop by 10.69\% and 9.12\%, respectively).
However, such results still represent the highest accuracy and the second-highest ranking in focal method coverage compared to baselines.

\subsection{Threats of Validity}
\noindent
\textbf{Internal Validity.}
The first one comes from potential data leakage since the referenced unit test written by developers may be part of the training data of LLMs (e.g., ChatGPT, Code Llama, etc.).
As for the close-source model, the precise training data remains inaccessible to us. 
The second one arises from potential errors in implementing our approach and baselines.
To minimize such a threat, we implement our model by pair programming and adopt the original source code of baselines from the ones shared by corresponding authors and use the same setting in the original papers. 
The authors also carefully review the experimental scripts to ensure their correctness.

\noindent
\textbf{External Validity.}
The main external threat to validity comes from our evaluation datasets used. 
The effectiveness observed in \toolname's performance may not be applicable across different unit test generation datasets, especially for unit tests written in other programming languages (e.g., C/C++, Python, etc.)

\begin{table*}[htbp]
  \centering
  \caption{The Results of cross-project demo selection strategy (Discussion)}
\resizebox{.95\linewidth}{!}{
    \begin{tabular}{l|l|r|r|rrr|rrr}
    \toprule
    Target   & \multicolumn{1}{l|}{Demo} & \multicolumn{1}{l|}{\% Size} & \multicolumn{1}{l|}{\# Query} & \multicolumn{1}{l}{Standard Acc.} & \multicolumn{1}{l}{Sample Acc.} & \multicolumn{1}{l|}{Decline} & \multicolumn{1}{l}{Standard fmc.} & \multicolumn{1}{l}{Sample fmc.} & \multicolumn{1}{l}{Decline} \\
    \midrule
    csv   & \multicolumn{1}{l|}{gson} & \multirow{2}[4]{*}{20\%} & 75    & 69.17\% & 57.33\% & -17.12\% & 77.42\% & 61.33\% & -20.78\% \\
    gson  & \multicolumn{1}{l|}{csv} &       & 43    & 63.18\% & 55.81\% & -11.67\% & 63.35\% & 58.14\% & -8.22\% \\
    chart & \multicolumn{1}{l|}{lang} & \multirow{2}[4]{*}{10\%} & 132   & 76.66\% & 67.67\% & -11.73\% & 82.55\% & 70.68\% & -14.38\% \\
    lang  & \multicolumn{1}{l|}{chart} &       & 272   & 84.13\% & 76.84\% & -8.67\% & 88.49\% & 84.93\% & -4.02\% \\
    \midrule
    Average &       &       & 522   & 78.37\% & 69.99\% & -10.69\% & 83.33\% & 75.73\% & -9.12\% \\
    \bottomrule
    \end{tabular}%
    }
  \label{tab:cross_pro_selection}%
\end{table*}%

\label{sec:threats}


\section{Related Work}

\subsection{Automated Testing}

The end-to-end unit testing solutions can be divided into two types: traditional approaches and large language model (LLM) based approaches. 
Traditional approaches include strategies based on search, randomization, or constraints-based, with the goal of maximizing test coverage (e.g., Evosuite~\cite{fraser2011evosuite}, Randoop~\cite{pacheco2007randoop}, and Agitar~\cite{2023agitar}).
However, these approaches have limited interpretability, especially the methods names and variables, which are hard to understand.
Recently, large language models have achieved big successes and can also be applied to software testing scenarios.
They can be further divided into two subgroups: approaches based on pre-training and fine-tuning paradigms (e.g., AthenaTest~\cite{tufano2020unit}, A3Test~\cite{alagarsamy2023a3test} and CAL-LM~\cite{rao2023cat}) and prompt-based testing generation (e.g., CodaMosa~\cite{lemieux2023codamosa}, QTypist~\cite{liu2023fill}, ChatUniTest~\cite{xie2023chatunitest} and TestPilot~\cite{schafer2023empirical}). 

\subsection{Prompt Engineering}

Prompt engineering refers to the methods of interacting with LLMs to guide their behavior and achieve desired outcomes without the need for updating the model weights~\cite{han2021pre,liu2023pre,sun2022paradigm}.
The few-shot learning~\cite{brown2020language,schick2020exploiting,schick2020s} and the instruction tuning~\cite{zhang2023instruction,reynolds2021prompt,mishra2021natural} are the most effective ways in prompt engineering.
Many studies~\cite{xie2021explanation,liu2021makes,webericl,lu2023survey} have explored how to select demonstrations to optimize performance and have observed that the choice of prompt format, the selection of examples of demonstrations, the number of the demonstrations, and the order of examples can lead to significant differences in performance. 
Providing examples is a way of implicitly describing tasks, while instruction tuning can directly convey human intent to LLMs. 
Instructing models with natural language commands can guide LLMs as precisely as possible. 
The studies~\cite{xie2021explanation,liu2021makes,lu2023survey} underscore the pivotal role of strategically selecting in-context examples and address the challenges of in-context and few-shot learning in large language models like GPT-3, highlighting the importance of example choice to enhance model generalization and consistency across diverse tasks. Collectively, these works~\cite{brown2020language,schick2020exploiting,schick2020s,xie2021explanation,liu2021makes,lu2023survey} contribute profound insights into understanding and optimizing the behaviors and capabilities of large pre-trained language models.

\label{sec:related_work}

\section{Conclusion}

This paper proposes a cascaded, model-agnostic, and end-to-end unit test generation framework.
\toolname first treats the unit test generation task as two cascaded tasks: test prefix generation and test oracle generation.
Then, to better stimulate models' learning ability, this paper manually builds large-scale demo pools to provide \toolname with high-quality test prefixes and test oracles examples.
Finally, \toolname automatically assembles the generated test prefixes and test oracles and compiles or executes them to check their effectiveness, optionally appending with several attempts to fix the errors outputted by the JUnit framework. 
The experimental results show the effectiveness and priority of \toolname over the studied three baselines.

\label{sec:conclude}


\vspace{-0.3cm}
\section*{Acknowledgements}
This work was supported by the National Natural Science Foundation of China (Grant No.62202419 and No. 62172214),
the Fundamental Research Funds for the Central Universities (No. 226-2022-00064), 
Zhejiang Provincial Natural Science Foundation of China (No. LY24F020008),
the Ningbo Natural Science Foundation (No. 2022J184), 
the Key Research and Development Program of Zhejiang Province (No.2021C01105), 
and the State Street Zhejiang University Technology Center.

\balance
\bibliographystyle{IEEEtran}
\bibliography{main}
\end{document}